\documentclass[journal]{IEEEtran}\twocolumn
\usepackage{amssymb,amsmath,cases,amstext}
\usepackage{graphicx}
\usepackage{algorithmic,algorithm,indentfirst}
\usepackage{setspace}
\usepackage{bm}
\usepackage{multirow}
\usepackage{rotating,enumerate}
\usepackage[T1]{fontenc}
\usepackage{lmodern}
\usepackage{color,framed}
\usepackage{subfig}
\usepackage{soul}
\usepackage{gensymb}
\definecolor{shadecolor}{rgb}{1,0.8,0.7}
\usepackage[nolist]{acronym}

\begin{document}

\title{Optimization of Demand Hotspot Capacities using Switched Multi-Element Antenna Equipped Small Cells}
\author{\IEEEauthorblockN{Hamed Ahmadi\IEEEauthorrefmark{1}, Danny Finn\IEEEauthorrefmark{1},
Rouzbeh Razavi\IEEEauthorrefmark{2}, Holger Claussen\IEEEauthorrefmark{2}, Luiz A. DaSilva\IEEEauthorrefmark{1}}\\
\IEEEauthorblockA{\IEEEauthorrefmark{1}{CTVR Telecommunications Research Center, Trinity College
Dublin, Ireland} \\
\IEEEauthorblockA{\IEEEauthorrefmark{2}Bell Laboratories, Alcatel-Lucent, Ireland}} \\
Email: {ahmadih,finnda,dasilval}@tcd.ie, {rouzbeh.razavi,holger.claussen}@alcatel-lucent.com \thanks{This material is based upon works
supported by the Science Foundation Ireland under Grants No. 10/IN.1/I3007, 10/CE/I1853 and 13/RC/2077.}}


\maketitle
\begin{abstract}
This paper presents switched Multi-Element Antennas (MEAs) as a simple, yet effective, method of enhancing the performance of small cell
heterogeneous networks and compensating for the small cell base station sub-optimal placement. The switched MEA system is a low-cost system
which enables the small cell to dynamically direct its transmission power toward locations of high user density, in other words
\emph{demand hotspots}. Our simulation results show that small cell base stations equipped with switched MEA systems offer greater
performance than base stations equipped with omni-directional antennas in terms of both the number of users that can be served (and hence
offloaded from the macrocell network) and in terms of overall network capacity. We also compare the performance of the switched MEA with
fixed directional antennas and show that fixed-directional antennas can only outperform the switched MEA if the misalignment between their
direction of transmission and the direction to the demand hotspot is less than $22.5^{\circ}$.
\end{abstract}




\begin{acronym}
%
%
%
%
%
\acro{3d}[3D]{3 Dimensional}
\acro{3gpp}[3GPP]{3\textsuperscript{rd} Generation Partnership Program}
\acro{aaa}[AAA]{Active Antenna Array}
\acro{abs}[ABS]{Almost Blank Subframe}
\acro{arq}[ARQ]{Automatic Repeat ReQuest}
\acro{awgn}[AWGN]{Additive White Gaussian Noise}
\acro{bcqi}[B-CQI]{Best \acs{cqi}}
\acro{bd}[BD]{Block Diagonalisation}
\acro{bicm}[BICM]{Bit-Interleaved Coded Modulation}
\acro{bler}[BLER]{BLock Error Rate}
\acro{ca}[CA]{Carrier Aggregation}
\acro{cbs-comp}[CBS-CoMP]{Coordinated Beam-Switching}
\acro{cdi}[CDI]{Channel Direction Indicator}
\acro{cdf}[CDF]{Cumulative Distribution Function}
\acro{ceuse}[CEUSE]{Cell Edge User Spectral Efficiency}
\acro{clsm}[CLSM]{Closed-Loop Spatial Multiplexing}
\acro{comp}[CoMP]{Coordinated Multi-Point}
\acro{cp}[CP]{Centralised Processor/coordinator}
\acro{cqi}[CQI]{Channel Quality Indicator}
\acro{cran}[C-RAN]{Cloud Radio Access Network}
\acro{crs}[CRS]{Common Reference Signal}
\acro{cs-comp}[CS-CoMP]{Coordinated Scheduling}
\acro{cs/cb}[CS/CB]{Coordinated Scheduling/Coordinated Beamforming}
\acro{cse}[CSE]{Cell Spectral Efficiency}
\acro{csi}[CSI]{Channel State Information}
\acro{csi-rs}[CSI-RS]{\acs{csi} Reference Signal}
\acro{csg}[CSG]{Closed Subscriber Group}
\acro{cs-rs}[CS-RS]{Cell Specific Reference Signal}
\acro{dcs}[DCS]{Dynamic Cell(/Point) Selection}
\acro{dm-rs}[DM-RS]{Demodulation Reference Signal}
\acro{dmt}[DMT]{Diversity-Multiplexing Tradeoff}
\acro{dpc}[DPC]{Dirty Paper Coding}
\acro{dps}[DPS]{Dynamic Point Selection}
\acro{dr}[DR]{Deployment Ratio}
\acro{dsl}[DSL]{Digital Subscriber Line}
\acro{e3f}[E$^3$F]{Energy Efficiency Evaluation Framework}
\acro{earth}[EARTH]{Energy Aware Radio and neTwork tecHnologies}
\acro{edn}[EDN]{Extremely Dense Network}
\acro{eesm}[EESM]{Exponential \acs{esm}}
\acro{enb}[eNB]{evolved Node Base station} 
\acro{epc}[EPC]{Evolved Packet Core}
\acro{esm}[ESM]{Effective \acs{sinr} Mapping}
\acro{fcs}[FCS]{Fast Cell Selection}
\acro{fdd}[FDD]{Frequency Division Duplexing}
\acro{ffr}[FFR]{Frequency Fractional Reuse}
\acro{fp7}[FP7]{European Commission 7\textsuperscript{th} Framework Programme}
\acro{gus}[GUS]{Greedy User Selection}
\acro{harq}[H-ARQ]{Hybrid Automatic Repeat reQuest}
\acro{henb}[HeNB]{Home/indoor femtocell \acs{enb}}
\acro{hspa}[HSPA]{High Speed Packet Access}
\acro{icic}[ICIC]{Inter-Cell Interference Coordination}
\acro{imt-a}[IMT-A]{International Mobile Telecommunications-Advanced}
\acro{inh}[InH]{Indoor Hotspot}
\acro{irc}[IRC]{Interference Rejection Combining}
\acro{jp}[JP]{Joint Processing}
\acro{jt}[JT]{Joint Transmission}
\acro{l2s}[L2S]{Link-to-System}
\acro{lesm}[LESM]{Logarithmic \acs{esm}}
\acro{ll}[LL]{Link Level}
\acro{los}[LOS]{Line-Of-Sight}
\acro{lte}[LTE]{Long Term Evolution}
\acro{lte-a}[LTE-A]{\acs{lte}-Advanced}
\acro{mac}[MAC]{Medium Access Control}
\acro{mcs}[MCS]{Modulation and Coding Scheme}
\acro{mea}[MEA]{Multi-Element Antenna}
\acro{mesc}[MESC]{Maximum Expected \acs{sinr} Combiner}
\acro{miesm}[MIESM]{Mutual Information Effective \acs{sinr} Mapping}
\acro{miESM}[MIESM]{Mutual Information \acs{esm}}
\acro{mimo}[MIMO]{Multiple Input Multiple Output}
\acro{ml}[ML]{Maximum Likelihood}
\acro{mmse}[MMSE]{Minimum Mean Square Error}
\acro{mno}[MNO]{Mobile Network Operator}
\acro{mrc}[MRC]{Maximum Ratio Combining}
\acro{mu-mimo}[MU-MIMO]{Multi-User \acs{mimo}}
\acro{mui}[MUI]{Multi-User Interference}
\acro{nas}[NAS]{Non-Access Stratum}
\acro{nl}[NL]{Network Level}
\acro{oda}[ODA]{Omni-Directional Antenna}
\acro{olsm}[OLSM]{Open-Loop Spatial Multiplexing}
\acro{oop}[OOP]{Object Oriented Programing}
\acro{pdcp}[PDCP]{Packet Data Convergence Protocol}
\acro{pf}[PF]{Proportional Fair}
\acro{phy}[PHY]{Physical} 
\acro{pmi}[PMI]{Precoding Matrix Indicator}
\acro{pon}[PON]{Passive Optical Network}
\acro{qos}[QoS]{Quality of Service}
\acro{rb}[RB]{Resource Block}
\acro{ri}[RI]{Rank Indicator}
\acro{re}[RE]{Resource Element}
\acro{rf}[RF]{Radio Frequency}
\acro{rlc}[RLC]{Radio Link Control}
\acro{rma}[RMa]{Rural Macrocell}
\acro{roi}[ROI]{Region of Interest}
\acro{rr}[RR]{Round Robin}
\acro{rrc}[RRC]{Radio Resource Control}
\acro{rrh}[RRH]{Remote Radio Head}
\acro{rrm}[RRM]{Radio Resource Management}
\acro{rsrp}[RSRP]{Reference Signal Received Power}
\acro{rue}[RUE]{Reassigned \acs{ue}}
\acro{rx}[Rx]{Receive}
\acro{samurai}[SAMURAI]{Spectrum Aggregation and \acs{mu-mimo}: ReAl-world Impact}
\acro{scbs}[SCBS]{Small Cell Base Station}
\acro{sdr}[SDR]{Software Defined Networks}
\acro{snr}[SNR]{Signal to Noise Ratio}
\acro{son}[SON]{Self-Organising Networks}
\acro{sinr}[SINR]{Signal to Interference and Noise Ratio}
\acro{siso}[SISO]{Single Input Single Output}
\acro{sl}[SL]{System Level}
\acro{sic}[SIC]{Successive Interference Cancellation}
\acro{sma}[SMa]{Suburban Macrocell}
\acro{ssps}[SSPS]{Semi-Static Point Selection}
\acro{su-mimo}[SU-MIMO]{Single-User \acs{mimo} beamforming on a single spatial stream}
\acro{su-mimo2}[SU-MIMO]{Single-User \acs{mimo}}
\acro{sus}[SUS]{Semi-orthogonal User Selection}
\acro{tbs}[TBS]{Transport Block Size}
\acro{tm}[TM]{Transmission Mode}
\acro{tr}[TR]{Technical Report}
\acro{ts}[TS]{Technical Specification}
\acro{tti}[TTI]{Transmission Time Interval}
\acro{tu}[TU]{Typical Urban}
\acro{tue}[TUE]{Target \acs{ue}}
\acro{tx}[Tx]{Transmit}
\acro{txd}[TxD]{Transmit Diversity}
\acro{ue}[UE]{User Equipment}
\acro{ula}[ULA]{Uniform Linear Array}
\acro{uma}[UMa]{Urban Macrocell}
\acro{umi}[UMi]{Urban Microcell}
\acro{urs}[URS]{\acs{ue}-specific Reference Signal}
\acro{wise}[WiSE]{Wireless System Engineering}
\acro{xgpon}[XG-PON]{10-Gigabit-capable \acs{pon}}
\acro{zf}[ZF]{Zero-Forcing}
\acro{zfbf}[ZFBF]{Zero-Forcing Beam Forming}
\end{acronym}

\section{Introduction}
Video streaming, online gaming and other data hungry applications cause high data consumption by mobile users. To deliver
the required data rates and to satisfy the requested \ac{qos} mobile operators are developing small cells close to demand
hotspots \cite{andrews2012femtocells}. A \emph{demand hotspot} is a location where a large number of network users are
gathered. Bus or tram stops, shopping streets, and off-street food markets are examples of possible demand hotspots.
Deploying \acp{scbs} at these hotspots poses two main challenges: firstly, caused by requirements of the \acp{scbs}, and
secondly, due to the dynamic nature of demand hotspots.

An \ac{scbs} needs backhaul and power supply connectivity, neither of which will be available at all desired deployment
locations, especially as network densities increase. Although using solar powered energy efficient \acp{scbs} can solve the
energy problem to some extent, providing reliable backhaul is still a challenge. There are also challenges in site
acquisition (even as small as a lamp post), in addition to energy and backhaul challenges. As a result, the \ac{scbs} will
be deployed at the nearest location that satisfies all these conditions.
Ideally the \ac{scbs} should be placed at the center of the demand hotspot but due to the aforementioned challenges it is
likely that the \ac{scbs} is deployed in a place that is not the center of the demand hotspot or it might not even be inside
the demand hotspot. This sub-optimal placement of the \ac{scbs} can significantly degrade the improvement that we expect by
deploying an \ac{scbs}.
%

The locations and appearance of demand hotspots change over the time. For example a demand hotspot around an off-street food
market only exists during lunch hours and its location depends on the (potentially non-static) location of the food market
on the street. This makes deploying an \ac{scbs} at the center of the temporary demand hotspot inefficient.

In order to provide good service 
in cases where \ac{scbs} cannot be placed optimally to cover a given demand hotspot, we consider the use of beamforming to
increase the signal gain in the direction of the desired demand hotspot. Compared to the use of an \ac{oda} at \ac{scbs},
beamforming enables the \ac{scbs} to direct its beam to the demand hotspot to overcome the inefficiencies caused by its
sub-optimal placement out of the demand hotspot. Ordinarily two options for this are considered: the use of an \ac{aaa}, or
the use of a fixed-beam directional antenna, at the \ac{scbs}.

Through independently controlling the phase and amplitude of multiple active antenna elements, \acp{aaa} in correlated
channels can be used to steer signal gains and nulls in desired directions \cite{Ghosh2010}. This is achieved by directing
constructive and destructive interference from the \ac{aaa} transmissions. As beamforming using \acp{aaa} is performed
electronically it can be programmed to self-configure and dynamically adjusted to suit the situation. Unfortunately, the use
of multiple active elements requires the small cell device to possess multiple transceiver chains, which can dramatically
increase the cost and size of the device. Further, the use of \acp{aaa} is not always supported for legacy \ac{ue} devices,
owing to the complex feedback that the beamforming requires.

\begin{figure}
 \centering
\subfloat[\footnotesize ODA]{\includegraphics[width=0.47\columnwidth]{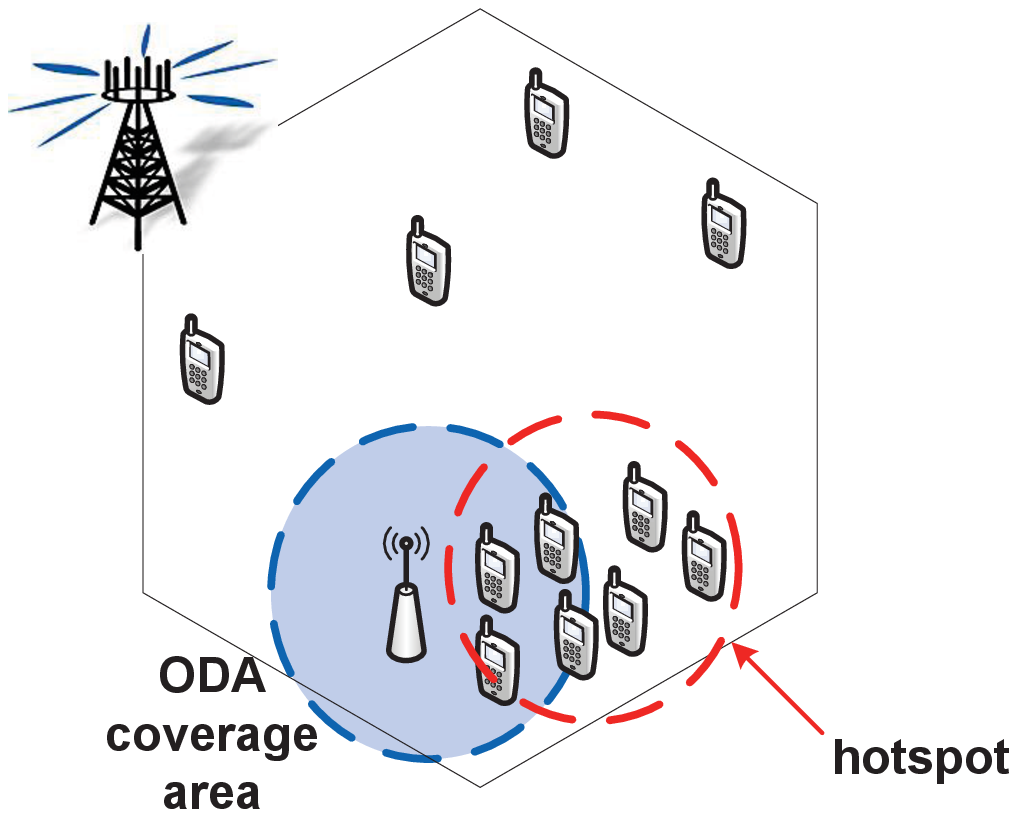}\label{fig:sysMa}} \subfloat[\footnotesize
Directional antenna]{\includegraphics[width=0.47\columnwidth]{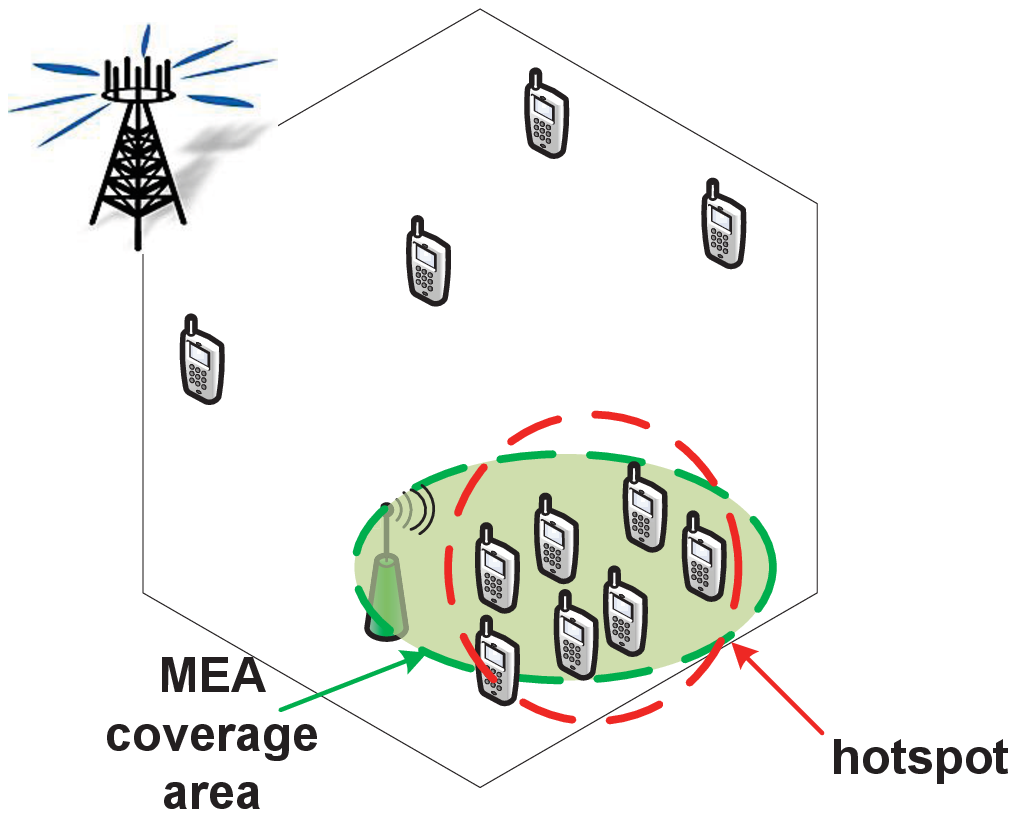}\label{fig:sysMb}}
  \caption{\footnotesize Small cell coverage with different antennas}\label{fig:sysModel}
\end{figure}
Alternatively, fixed-beam directional antennas can be used to provide \ac{scbs} beamforming gains in the direction of
activity hotspots \cite{calin2013high}. One such example is the use of double patch antennas which emit a focused beam.
While this presents a more cost effective and legacy \ac{ue} compatible solution, the direction of the fixed beam must be
configured manually at the time of \ac{scbs} installation and as such is sensitive to misconfiguration and cannot
dynamically adjust its operation to effectively serve demand hotspots. In Figure \label{fig:sysModel} we show how different
the coverage of \ac{scbs} would be when it is equipped with an omni-directional antenna and a directional antenna.

Another option, which is the focus of this work, is the use of switched \acp{mea} \cite{claussen2009femtocell,
Razavi_MEA2014}.
In this solution a beamforming gain is obtained in the desired direction by selecting, from multiple differently orientated
antenna elements, the one which provides the strongest gain in the demand hotspot direction. As only a single antenna
element transmits at each point in time, this solution requires only a single transceiver chain, meaning that it provides a
low cost beamforming solution which is capable of self-configuring. Further, by dynamically reselecting the antenna element
used for transmission the switched \acp{mea} is also capable of changing its beam direction if the location of the hotspot
changes. In \cite{claussen2009femtocell}, the switched \ac{mea} system is introduced as a low cost solution for residential
femtocells to increase the indoor coverage and reduce the number of mobility events using the mobility-event-based
self-optimizing approach \cite{Claussen08femtocover}.


In this paper we use switched \acp{mea} to direct \ac{scbs} transmissions toward the desired hotspots. We show that using
switched \acp{mea} firstly increases the number of UEs served by the small cell and secondly improves the system performance
in terms of total data rate. In this work we also investigate the antenna selection problem and the amount of required
samples for confidently selecting an antenna element.


\section{System model}

In this work we consider a scenario in which there are a number of macrocell base stations (BS) covering the considered area
and each BS has three sectors. User are randomly distributed in this area; however, there exist some demand hotspots. We
deploy small cells to improve the network performance in terms of providing higher data rate for the users. Ideally the
SCBSs are deployed at the center the hotspots, but as mentioned, in reality we might not be able to place an SCBS at the
center of each demand hotspot. Here, we consider SCBSs that are equipped with switched MEA system (MEA-SCBS).

\begin{figure}
 \centering
\subfloat[\footnotesize Single patch]{\includegraphics[width=0.47\columnwidth]{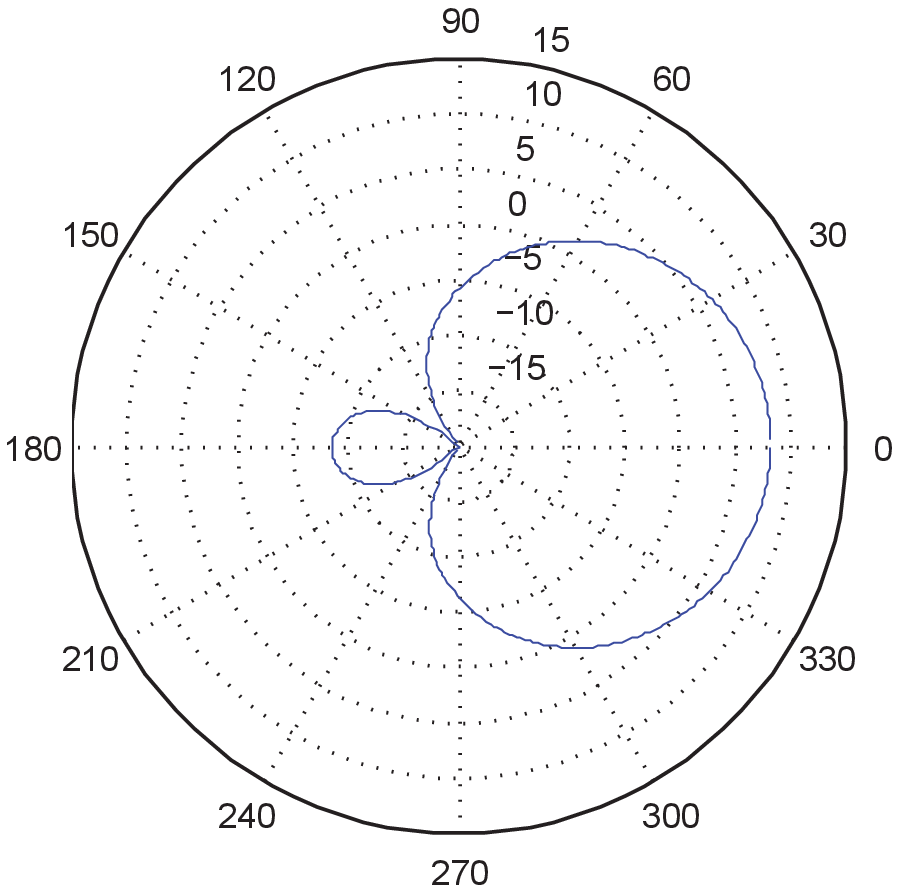}\label{fig:SPgain}}
\subfloat[\footnotesize Double patch]{\includegraphics[width=0.47\columnwidth]{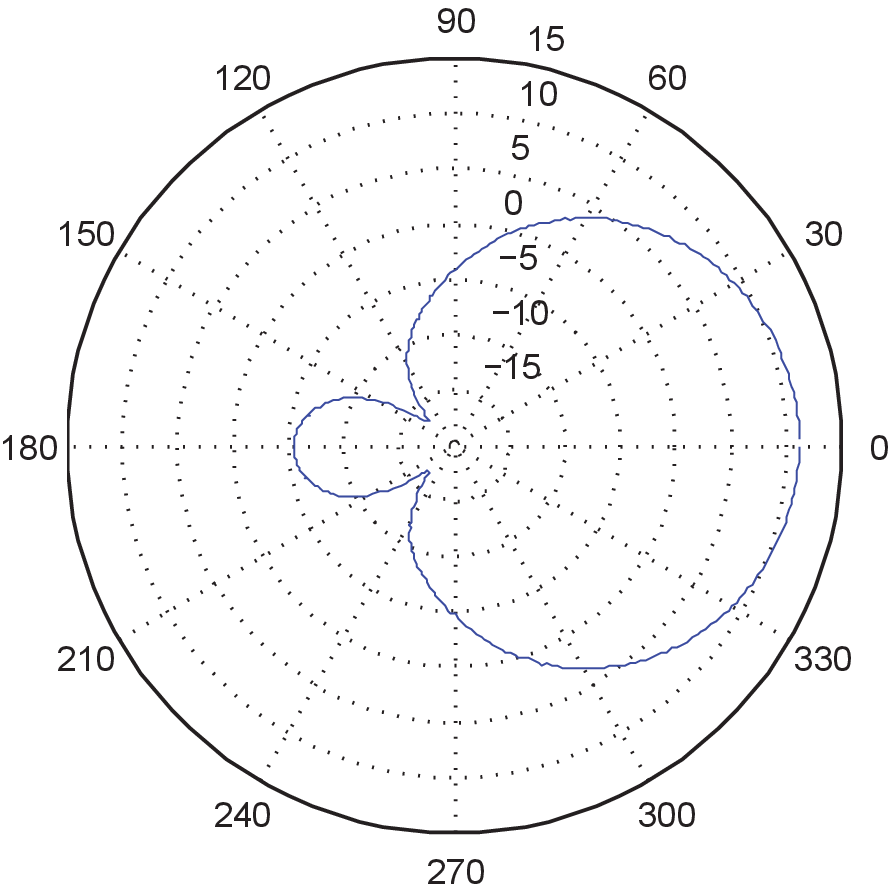}\label{fig:DPgain}}
  \caption{\footnotesize Antenna pattern}\label{fig:SPDPgain}
\end{figure}

A switched Multi-element antenna system is a simple solution to benefit from beam directionality. This can be created by
equipping the SCBS with multiple antenna elements with distinctive antenna patterns and a switch to select a single or a
combination of antenna elements. In this work we consider an MEA with four orthogonal antenna elements. The considered
antenna elements are all either single patch or double patch antennas. While the single patch antenna is lighter and smaller
in the size, the double patch antenna provides higher antenna gain. The antenna gain pattern of single patch and double
patch antennas are shown in Figures~\ref{fig:SPgain} and \ref{fig:DPgain}, respectively.

For simplicity we assume that the MEA-SCBSs can activate only one of their antenna elements. The MEA-SCBS selects its best
antenna element
that captures most of users in the demand hotspot. 

\section{Antenna selection}
The best antenna element of any MEA-SCBS can be centrally selected. In this approach the best antenna element is the one
that maximizes the system's throughput, which is shown as
\begin{equation}\label{eq1}
a^*= \text{argmax } r(a_i), i\in\{1,2,3,4\},
\end{equation}
where $a^*$ the best antenna and $r(a_i)$ is the total throughput when $i^{th}$ antenna element is selected at the
\ac{scbs}. The throughput is computed by the modified Shannon capacity formula \cite{Mogensen07_modShan}.

Using the centralized method imposes high computational complexity on the central system and also causes signalling
overhead. Distributed decision making is the alternative solution to this problem where each MEA-SCBS selects its best
antenna element. Considering that each UE will be served by the SCBS if the received signal power from the SCBS is more than
the received signal power from the macrocell base station. The number of UEs that can be served by each antenna element is a
metric that
%
can be evaluated based on local information. Therefore,
the MEA-SCBS can select its desired antenna element in a fully distributed manner.

To select the best antenna, the MEA-SCBS in a continues order turns each of its four elements on and saves the number of UEs
that each antenna element serves. The selected antenna can simply be the antenna that serves the most UEs.
\begin{equation}\label{eq2}
a^*= \text{argmax } s_{UE}(a_i), i\in\{1,2,3,4\},
\end{equation}
where $s_{UE}(a_i)$ gives the number of UEs served by the \ac{scbs} when $i^{th}$ antenna is active.

Distribution of UEs in the demand hotspot and the noises in the system can affect the accuracy of selecting the best
antenna. Therefore, it is possible that we need to perform the decision iterations multiple times and make the final
decision based on the majority rule. We assess the number of decision iterations required based on two definitions of the
best antenna: angle-based and T-test-based.

\subsection{Angle-based}
In this model we considered the true best antenna element to be the one for which the difference between the angle in which
the element is directed and the direction of the hotspot centre-point is smallest. The antenna selection will then be
performed based on which antenna element could serve the highest average number of UEs across multiple iterations, between
which the UE distribution changed. The the number of times that the antenna with the smallest angle has been selected shows
how accurate the majority rule is at that specific number of iterations.

\subsection{T-test-based}
Since one third of the UEs are contained in the hotspot area, the best antenna element should serve a significantly higher
number of UEs than the other antenna elements, regardless of the user distribution. However, cases in which the angle
between the center of the hotspot and the antenna direction is close to $45$ degrees are more challenging, i.e., two antenna
elements might serve a similar number of UEs. Therefore, we use a T-test to realize after how many iterations the selected
antenna element significantly outperforms the others.

\section{Simulation results}
We consider a standard 3GPP scenario where a tri-sector macrocell is surrounded by 6 other similar cells. We focus on the
performance of a single SCBS, attempting to serve a single demand hotspot, where both the SCBS and hotspot are randomly
dropped within the coverage region of a selected sector of the central macrocell.

Distributed throughout the macrocell are $30$ UEs: two thirds are dropped randomly while one third are concentrated within
the demand hotspot. The hotspot is considered to be a circular area with a radius of $10$ meters.

The relative locations of the demand hotspot and the SCBS are characterised by $\gamma_{HS}$ which we define as the received
SINR at the centre-point of the hotspot, where the signal source is the SCBS equipped with an ODA and the interference
sources are the surrounding macrocells. The macrocell maximum transmit power is $46$ dBm and the small cell transmit power
is $20$ dBm. All base stations have a bandwidth of 10 MHz and 3GPP outdoor scenario pathloss models are applied.


\subsection{Antenna selection training}
As mentioned, to select the best antenna the MEA-SCBS turns on each of its antenna elements in turn and checks how many UEs
it can serve with each of them. The antenna that can serve the highest average number of UEs after a certain number of such
iterations is then selected. In this subsection we investigate the affect of changes in the user distribution at the hotspot
on the antenna selection accuracy. In other words the number of rounds that the MEA-SCBS should check all the four antenna
elements before making a decision.

Assuming that during the decision period the locations of the SCBS and the hotspot do not change, the worst case scenario
that can be considered is that the distributions of UEs within the hotspot and the cell region change each time that the
SCBS checks the number of UEs served by each antenna. In this set of simulations we considered $1000$ random drops of the
hotspot and SCBS locations. For each drop and each decision round we considered a different UE distribution and investigated
its effect on the antenna selection.




\begin{figure}
    \centering
    \includegraphics[width=0.9\columnwidth]{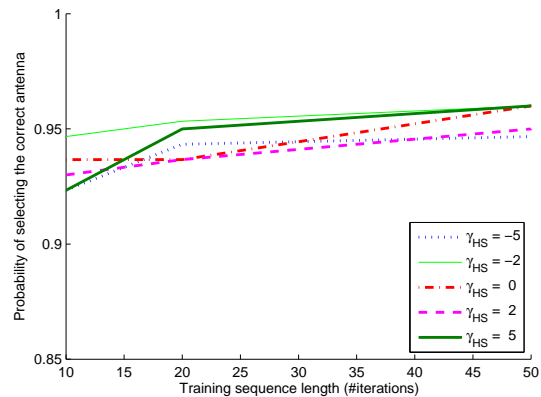}
    \caption{\footnotesize Best antenna is known: the probability of selecting the best antenna.}\label{fig:ClassLabel}
\end{figure}


\subsubsection{Angle-based} Figure~\ref{fig:ClassLabel} shows the relation between the number of rounds over which the antenna selection was averaged
(number of times that each antenna element was checked) and the probability of selecting the true best antenna. The figure
shows that correct selection of the best antenna is more probable with more samples; however, the increment is quite small,
and the probability of selecting the best antenna element with only $10$ rounds is over $0.9$. We also see that the
likelihood of correct antenna selection is not significantly affected by the SINR at the demand hotspot.




\subsubsection{T-test-based} Table~\ref{table1} shows the average number of rounds required for the best antenna element to pass the $95\%$ significance
T-test.
Similar to the previous test we observe that the required number of rounds is almost the same for different SINR values.

Even though, according to the table, on average almost $16$ rounds are needed to show that one antenna element significantly
outperforms all others, we also found that within \textit{all} simulated iterations, simply selecting the antenna element
which served the most UEs (based on a single iteration alone) also provided the same outcome. In other words, for a given
SCBS and hotspot placement, the antenna element that served the most UEs remained the same for all user distributions. For
this reason, in the remainder of this paper, selection of the best antenna by the MEA-SCBS is performed by comparing the
number of UEs served by each antenna element for only a single round.

\begin{table}
\begin{center}
\scalebox{0.85}{
\begin{tabular}{ |c|c|c|c|c|c| }
 \hline
 ODA SINR at hotspot center  & $-5$ & $-2$ & $0$& $2$ & $5$\\
 \hline
 Training sequence length & $15.1$ & $16$ & $14.7$ & $15.5$ & $15.2$ \\
 \hline

\end{tabular}
}
\end{center}\caption{\footnotesize T-test-based method: required training to confidently select the antenna.}\label{table1}
\end{table}

\subsection{Number of served UEs}
In this subsection we compare the performance of the switched MEA-equipped SCBS to ODA-equipped and fixed-direction SCBSs.
We compare this in terms of the number of UEs that the SCBS can serve.

For the fixed-direction SCBSs we consider that, as the location of the hotspot may not be accurately known at the
installation time, there exists some probability of misalignment. For this reason we consider possible misalignments of
20$^{\circ}$, 40$^{\circ}$ and 60$^{\circ}$, as well as the ideal alignment case.

\begin{figure}
\centering
\includegraphics[width=0.9\columnwidth]{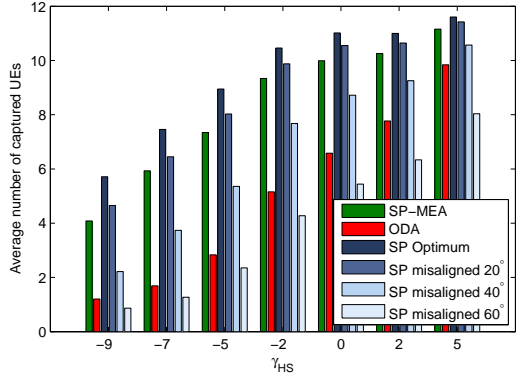}
\caption{\footnotesize Average number of served UEs for different antenna configurations. Here a single patch directional
antenna is considered for the cases of optimally directed at the hotspot, misaligned by 20$^{\circ}$, 40$^{\circ}$, and
60$^{\circ}$, and the switched MEA case.}\label{fig:UEnoMiss20}
\end{figure}

Figure~\ref{fig:UEnoMiss20} compares the number of served UEs by an SCBS equipped with an ODA, a switched MEA and
fixed-direction antenna, where single-patch direction antennas are used. The switched multi-element antenna outperforms the
omni-directional antenna which is more significant for lower values of $\gamma_{HS}$. The figure also illustrates that the
fixed-directional antennas outperform the switched MEA only if the misalignment is less than roughly $25$ degrees.

\begin{figure}
\centering
\includegraphics[width=0.9\columnwidth]{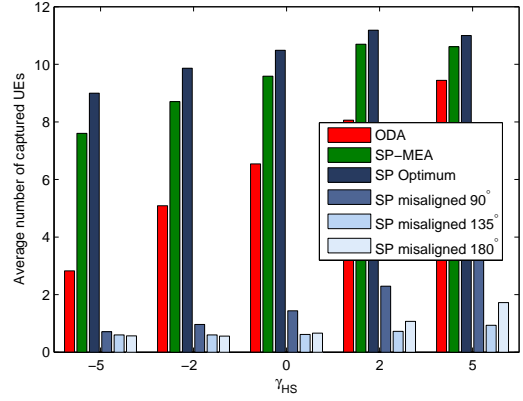}
\caption{\footnotesize Average number of served UEs for different antenna configurations when the fixed-directional
misalignment is large. Here a single patch directional antenna is considered for the cases where the directed antenna is
misaligned with the hotspot direction by 90$^{\circ}$, 135$^{\circ}$, and 180$^{\circ}$, as well as the optimally aligned
and switched MEA cases.}\label{fig:UEnoMiss90}
\end{figure}

The hotspot location may change in time and therefore there are cases in which the installed directional antenna may end up
pointing towards a completely different direction. In Figure~\ref{fig:UEnoMiss90}, we investigated this scenario. The figure
clearly shows that in such situations the fixed-directional antennas have poor performance, while the MEA and ODA cases
remain unaffected by the major misalignment.

\begin{figure}
\centering
\includegraphics[width=0.9\columnwidth]{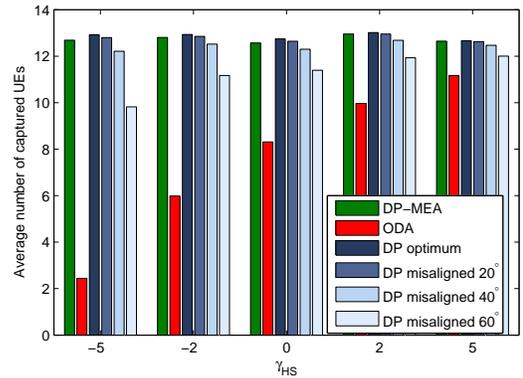}
\caption{\footnotesize Average number of served UEs for different antenna configurations. Here a double patch directional
antenna is considered for the cases of optimally directed at the hotspot, misaligned by 20$^{\circ}$, 40$^{\circ}$, and
60$^{\circ}$, and the switched MEA case.}\label{fig:UEnoMiss20DP}
\end{figure}

As mentioned previously the antenna gain of the directional antenna can be increased by using two joint antenna patches
(double patch) instead of a single antenna patch. The same model can be used in the multi-element antennas and as a result
the MEA will have a double-patch antenna in each of its directions which improves its performance.
Figure~\ref{fig:UEnoMiss20DP} presents the number of UEs that the MEA and directional antennas with different misalignments
can serve using double-patch antennas. The figure shows that while the omni-directional antenna's performance decreases
significantly with $\gamma_{HS}$, the double-patch MEA and directional antennas are still capable of covering almost all UEs
present within the hotspot area. This figure also illustrates that the effect of misalignment is not as significant in the
double patch case as it is for single-patch directional antennas. However, the misalignment is more apparent in lower SINRs.

\subsection{System performance}

\begin{figure}
\centering
\includegraphics[width=0.9\columnwidth]{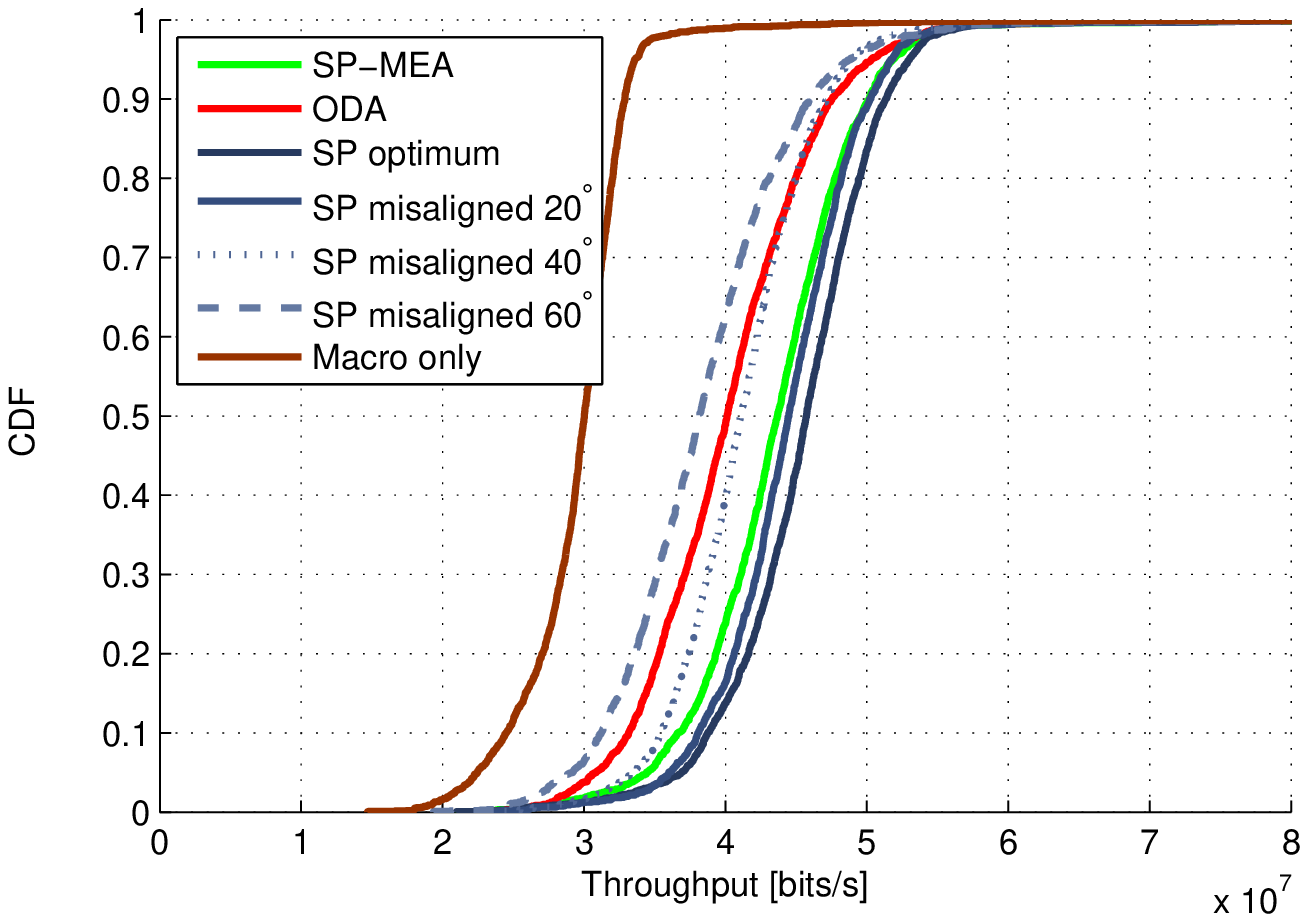}
\caption{\footnotesize System performance using different antennas at the SCBS, when the received SINR from the ODA is 0 dB
at the center of the hotspot.}\label{fig:SPsysrate}
\end{figure}

\begin{figure}
\centering
\includegraphics[width=0.9\columnwidth]{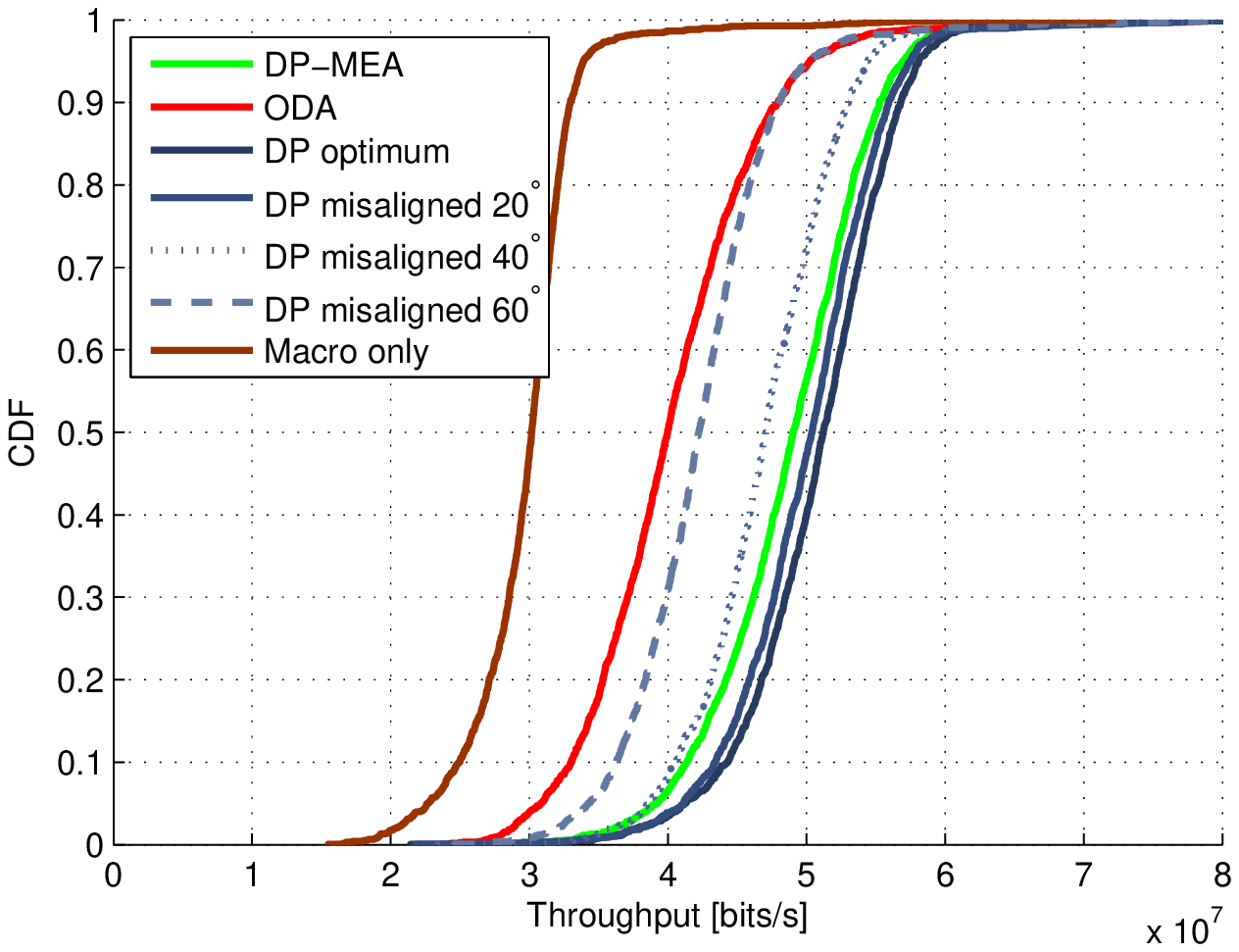}
\caption{\footnotesize System performance using different antennas at the SCBS, when the received SINR from the ODA is 0 dB
at the center of the hotspot.}\label{fig:DPsysrate}
\end{figure}

As the small cell base stations are deployed to assist the macrocell in providing UE higher data rates, we measure the
performance of the system in terms of the total data rate. Figure~\ref{fig:SPsysrate} shows the CDF plot of UEs' data rates
in the system. The performance of the system without a small cell (macrocell only) is also shown in the figure. The figure
clearly shows that using the switched MEA improves the data rate for UEs in the system. The same scenario with the use of
double-patch antennas is presented in Figure~\ref{fig:DPsysrate}. The figures show that the single-patch MEA and
double-patch MEA increase the system rate compared to the omni-directional antenna-based small cell by $10\%$ and $25\%$,
respectively.

%
\section{Conclusions}
In this paper we investigated the performance of switched \acp{mea} as a simple solution to enable the small cells to direct
their transmit power. Our simulation results showed that the switched \ac{mea} system can accurately train itself and for
the training it does not require any additional information. The switched \ac{mea} was able to serve more UEs than the
omnidirectional antenna and the difference was more significant where the switched \ac{mea} had double patch antenna
elements. We also show that using switched \acp{mea} instead of omnidirectional antennas can improve the systems data rate
up to $25\%$. It is important to note that unlike the fixed directional antennas the switched \acp{mea} are flexible and
they can change their beam direction when the location of the demand hotspot changes.

In our previous work \cite{Razavi_MEA2014} we provided a techno-economical analysis and discussed the cost efficiency of the
switched \ac{mea} system. Our analysis in this work showed that using multi-element enables the operators to install the
SCBS in the locations that are further from the center of the demand hotspot and still achieve the same performance as they
used to achieve. This may help the operators to save site rental costs.

To extend our findings in our future works, we will study the interactions among multiple switched MEA-SCBSs. The switched
MEA-SCBSs cooperate to improve the systems data rate by avoiding interfering each other. The MEAs can either serve UEs of
multiple hotspots or collaboratively serve UEs of a single demand hotspot. We will also study the self-healing mechanisms in
such a system.


\bibliographystyle{ieeetr}
\bibliography{ALU}

\end{document}